\documentclass[twocolumn,natbib]{svjour3}
\usepackage{mathptmx}
\usepackage[T1]{fontenc}
\usepackage[utf8]{inputenc}
\usepackage{array}
\usepackage{verbatim}
\usepackage{fancybox}
\usepackage{calc}
\usepackage{marvosym}
\usepackage{url}
\usepackage{multirow}
\usepackage{amstext}
\usepackage{graphicx}
\usepackage{booktabs}
\usepackage{tabularx}
\usepackage{subfigure}
\usepackage{color}
\usepackage{textcomp}

\makeatletter

\RequirePackage{fix-cm}

\smartqed  

\usepackage{marvosym}
\usepackage{epstopdf}
\usepackage{SIunits}
\usepackage {ulem}

\newcommand{\graph}[1]{\centerline{
\includegraphics[width=1\columnwidth]{#1}
}}
\newcommand{\sect}[1]{Section~\ref{#1}}
\newcommand{\fig}[1]{Figure~\ref{#1}}

\def\mum{\nobreak\mbox{$\;$\textnormal{\textmu m}}}

\journalname{Rheologica Acta}

\makeatother

\begin{document}

\title{Particles accelerate the detachment of viscous liquids
}


\author{Merlijn S. van Deen \and Thibault Bertrand \and Nhung Vu \and David Quéré \and Eric Clément \and Anke Lindner}

\institute{Merlijn S. van Deen \and Thibault Bertrand \and Nhung Vu \and David Quéré \and Eric Clément \and Anke Lindner (\Letter{})\at
PMMH -- ESPCI, 10, rue Vauquelin, 75231 Paris Cedex 05, France \\
\email{anke.lindner@espci.fr}
\and Nhung Vu \and David Quéré \at LadHyx, École Polytechnique, 91120 Palaiseau, France
\and Merlijn S. van Deen \at Kamerlingh
Onnes Lab, Universiteit Leiden, Postbus 9504, 2300 RA Leiden, The
Netherlands
\and Thibault Bertrand (present address) \at
{Department of Mechanical Engineering and Material Sciences \\
Yale University, New Haven, Connecticut 06520-8286, USA}}

\date{Received: date / Accepted: date}
\maketitle
\begin{abstract}
During detachment of a viscous fluid extruded from a nozzle a filament linking the droplet to the latter is formed. Under the effect of surface tension the filament thins until pinch off and final detachment of the droplet. In this paper we study the effect of the presence of individual particles trapped in the filament on the detachment dynamics using granular suspensions of small volume fractions ($\phi<6\%$). We show that even a single particle strongly modifies the detachment dynamics. The particle perturbs the thinning of the thread and a large droplet of fluid around the particle is formed. This perturbation leads to an acceleration of the detachment of the droplet compared to the detachment observed for a pure fluid. We quantify this acceleration for single particles of different sizes and link it to similar observations for suspensions of small volume fractions.  Our study also gives more insight into particulate effects on detachment of more dense suspensions and allows to explain the accelerated detachment close to final pinch off observed previously \citep{Bonnoit2010Accelerated}.

\PACS{ 
 83.50.Jf 
 \and 83.80.Hj 
 \and 83.85.Ei 
 \and 47.55.db 
 \and 47.57.E- 
} %
\keywords{Granular suspensions \and Extensional flow \and Drop formation}
\end{abstract}

\section{Introduction}
\label{sec:intro}

At the end of the 19th century, Plateau and Rayleigh had devised a description for the breakup of jets in droplets. They described the start of the breakup, but could not yet investigate the behavior close to detachment, or 'pinch-off', which happens at very short time scales and small length scales. The broad adoption of tools such as high-speed cameras has allowed research into this behavior close to the detachment. A large number of these recent theoretical and experimental papers was reviewed by \cite{Eggers_Nonlineardynamicsand_RoMP1997} and the detachment dynamics for simple liquids are by now well understood.

This has also initiated the use of capillary breakup as a
rheological technique, allowing to directly access the viscosity of
a given fluid \citep{McKinley2000}. The thinning of the filament,
leading to "pinch-off" implies a strong elongation and thus
allows to study viscous properties under this kind of deformation. Experiments
with polymer solutions have shown that this experimental situation
is indeed a good set-up to access elongational properties of these
fluids \citep{Amarouchene2001, Tirtaatmadja2006, Sattler2008, anna2001elastocapillarythinning}. The stability of jets formed by foams, pastes, polymeric liquids, dry grains, or grains
immersed in a surrounding liquid has also been addressed
\citep{Pignatel2009, Lohse2004, Royer2009, Lespiat2010, Coussot2005, clasen2012dispensing, christanti2001surfacetensiondriven, christanti2002effectfluidrelaxation}.

Lately the detachment of particle laden liquids has received a lot of interest \citep{furbank2004experimental, furbank2007pendant, Bonnoit2010Accelerated, Bertrand2011Dynamics, Miskin2011DropletA, Roche2011Heterogeneity}, partially due to the fact that this is of importance for industrial processes as food processing or inkjet printing \citep{Morrison2010}. The focus of these studies was twofold. On the one hand, the droplet detachment yields access to effective properties of the suspension under elongation and on the other hand particulate effects on the stability of the jet and the droplet detachment can be investigated. Several studies have shown that the detachment of a granular suspension can initially be described by its effective viscosity \citep{furbank2007pendant, Bonnoit2010Accelerated, Bertrand2011Dynamics} and the elongational viscosity is thus simply given by three times the shear viscosity.   At later stages of the detachment the dynamics of the particles become important and lead to very specific detachment dynamics \citep{furbank2004experimental, furbank2007pendant, Bonnoit2010Accelerated, Bertrand2011Dynamics, Miskin2011DropletA}.
It has been shown that this behavior can be described as an accelerated detachment \citep{furbank2007pendant, Bonnoit2010Accelerated}, where the overall acceleration depends on both particle size and volume fraction \citep{Bonnoit2010Accelerated}. The origin of the specific dynamics observed during the detachment, in particular of dense suspensions, is still under debate \citep{fall2012granular} and we will focus in this paper on a simpler system.

Following the pioneering work of \cite{furbank2004experimental, furbank2007pendant} we will in this paper focus on small volume fractions ($1-6\%$). For these dilute suspensions, the effective viscosity of the suspension remains nearly identical to the viscosity of the suspending fluid and the role of individual particles in the neck of the droplet on the detachment dynamics can be addressed unambiguously. We will show that the addition of a very small number of particles significantly changes the detachment dynamics and leads to an acceleration of the final stages of the detachment compared to the pure fluid. We will also show that this study helps to investigate the thinning behavior of denser suspensions.

In \sect{sec:background} we introduce droplet detachment of simple fluids and our earlier work on model suspensions. In \sect{sec:setup} we describe the experimental setup and the suspensions we have used. We describe our measurements on the behavior of suspensions with volume fractions up to 6\% in \sect{sec:low phi}. In \sect{sec:single particle}, we show the importance of individual particles in the neck. In \sect{sec:PIV}, we discuss the link between out study and the detachment of more dense suspensions. In \sect{sec:conclusion} we conclude.

\section{Droplet detachment of fluids and suspensions}
\label{sec:background}

\begin{figure}
\graph{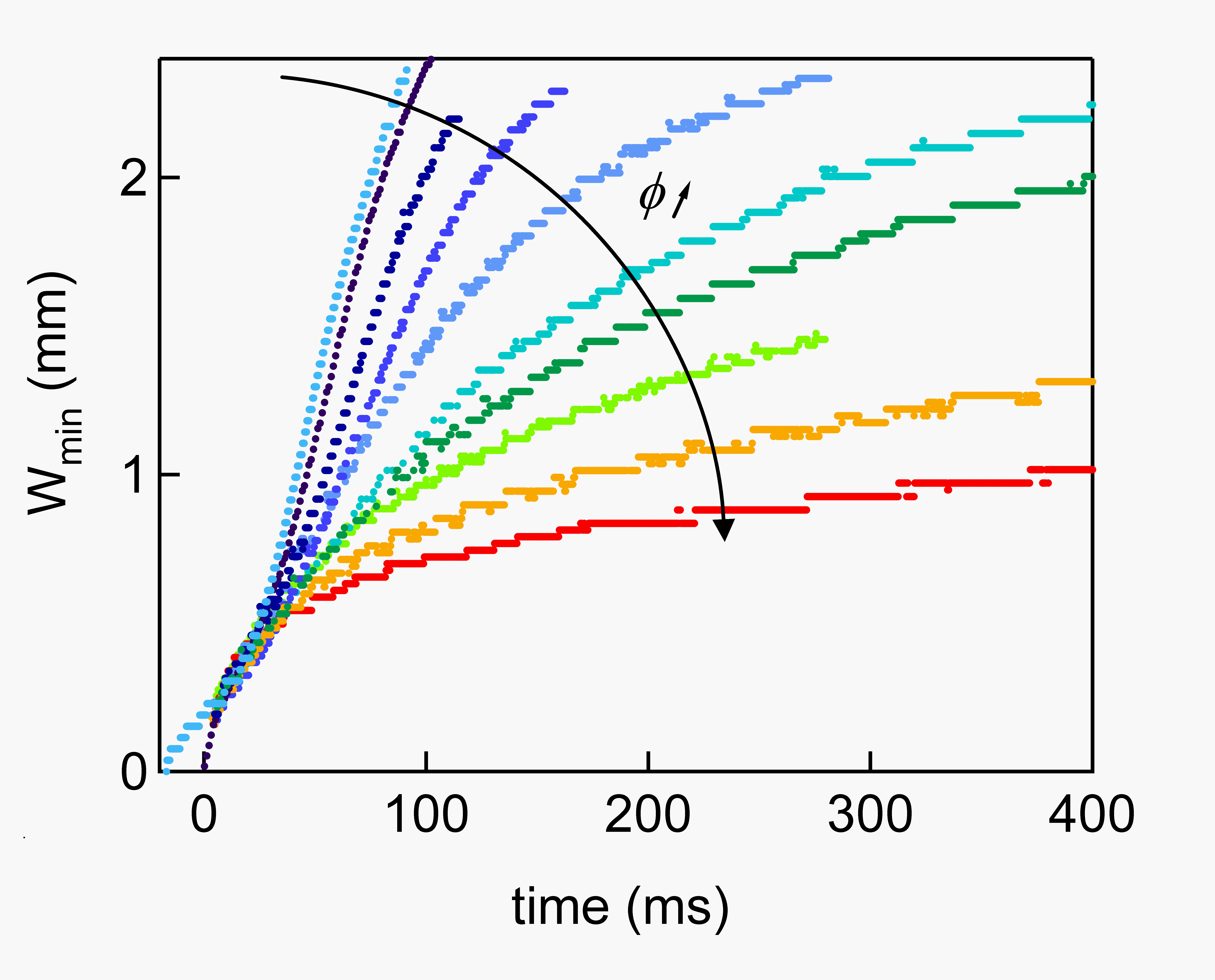}
\caption{Time evolution of the minimum neck diameter $W$ for suspensions
    of volume fractions $\phi=15$,~$20$,~$30$,~$40$,~$45$,~$48$,~$50$,~$ 53$,~$ 55\%$
    with grain diameter $d=40\mum$. The most left curve  (cyan) corresponds to pure interstitial fluid ($\phi=0$). The origin of the $x$-axis is given by the time of the pinch for each suspension, the curve for the interstitial fluid is shifted by $\Delta t=-20$ ms compared to its pinch time (data from \cite{Bonnoit2010Accelerated}). \label{fig_regimes_POF}}
\end{figure}

\subsection{Droplet detachment of viscous Newtonian fluids}
Once gravity overcomes the surface tension of a droplet hanging from a syringe, the droplet starts falling. At this point the neck, which connects the bulk of the droplet to the syringe, starts to thin. This behavior is not, as one might expect, governed by gravity, but by surface tension: the surface tension, which was holding the droplet together before it fell, is now the driving force.

The initiation of the thinning is due to the Rayleigh-Plateau instability leading to an exponential decrease of the neck radius in time.
Once the neck is sufficiently thin, the detachment transitions into a second, self-similar regime. The behavior in this regime is independent of initial conditions, and scales for viscous fluids linearly with the time to detachment. This regime can be divided into two sub-regimes: the earlier \cite{papageorgiou1995breakup} or Stokes regime and the later \cite{eggers1993universal} or Navier-Stokes regime. \cite{rothert2003formation} have experimentally shown the existence of these regimes, and have described the transitions between the three regimes. Because the transition to the Eggers regime happens at small neck radii, the effect of this regime on the breakup time can be neglected for not too viscous fluids \citep{clasen2012dispensing, rothert2003formation, basaran2002smallscalefree}.

\subsection{Droplet detachment of suspensions -- previous work and current study}

In earlier work from our group we have described the detachment of droplets of density matched, non-Brownian suspensions as taking place via different detachment regimes \citep{Bonnoit2010Accelerated,Bertrand2011Dynamics}. We showed that the effective shear viscosity governs the droplet detachment of suspensions far from the pinch-off: a pure fluid and a suspension with the same effective viscosity show the same thinning behavior. As the thinning goes on particles rearrange in the thread and areas less dense in grains are created. One observes a crossover from the initial \emph{effective fluid regime} towards an \emph{interstitial fluid regime}, where the dynamics are governed purely by the interstitial fluid. Finally, the droplet detaches quicker than one would expect for the interstitial oil, the \emph{accelerated regime}.

\begin{figure}
\begin{center}
\subfigure[Experimental set-up \label{fig:set-up}]{\includegraphics[width=0.95\columnwidth]{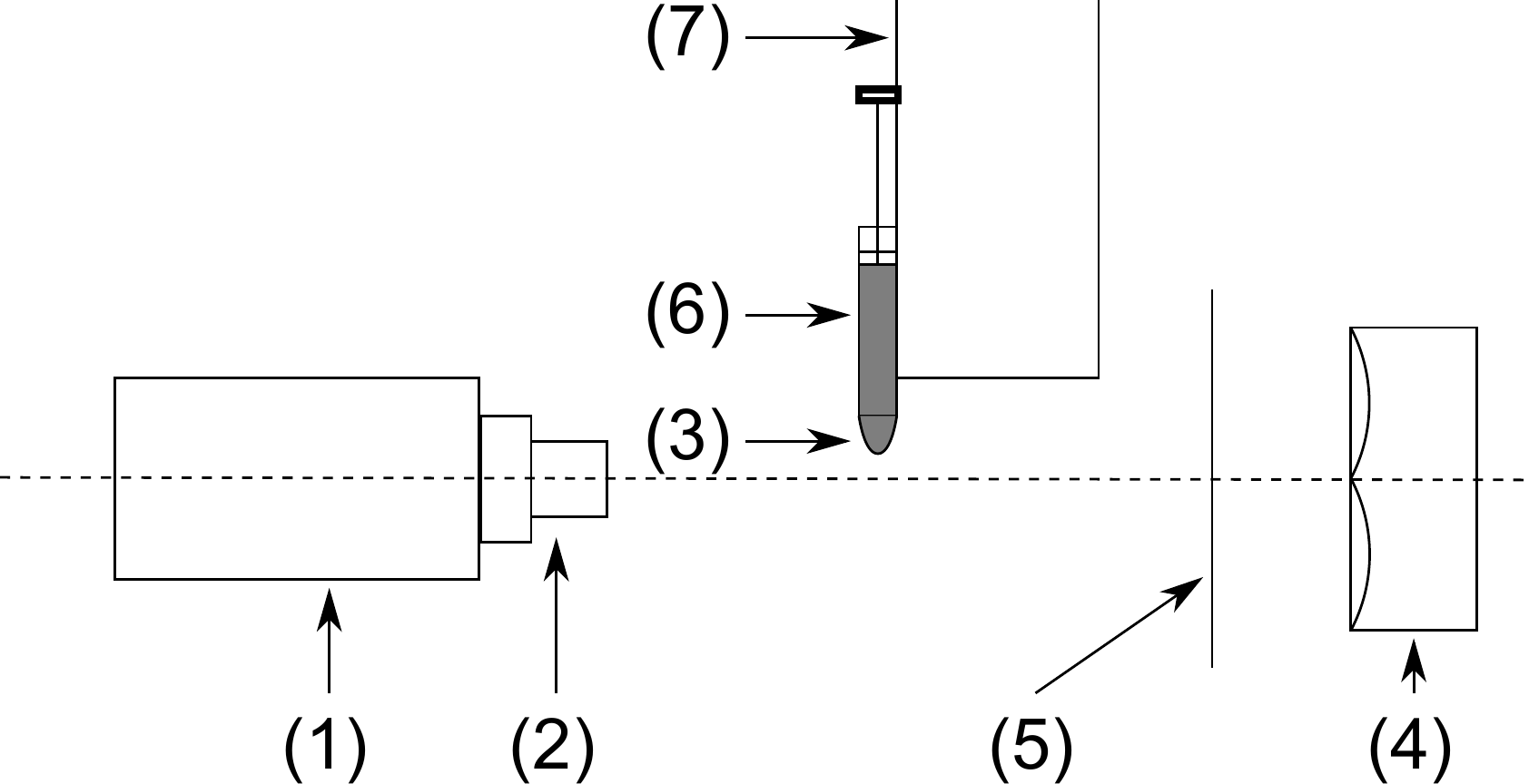}}
\hfill
\subfigure[Typical detachment of a low volume fraction granular suspension. \label{fig:detach_susp}]{\includegraphics[width=0.8\columnwidth]{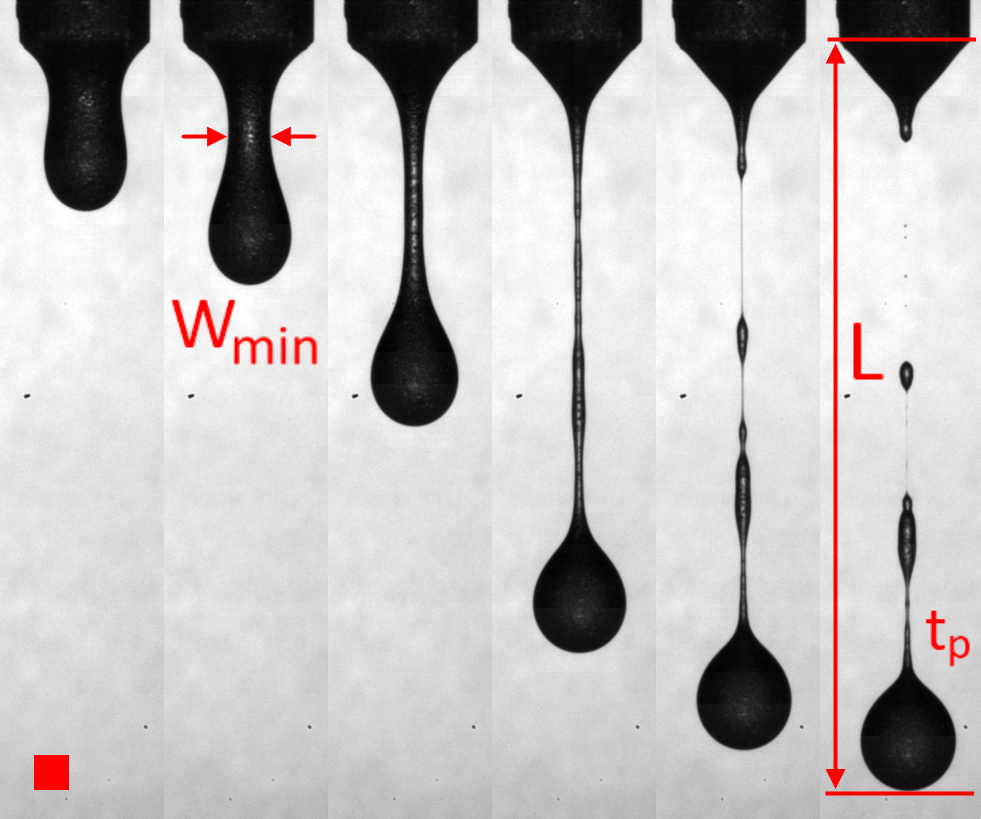}}
\caption{(a) The setup used in our experiments. Using a high speed camera (1) with
a macro lens (2), we image the droplet (3), which is back - lighted using a high-power LED spot (4) with diffuser (5). The droplet is extruded from a syringe (6)
using a syringe pump (7) at a low flow rate. (b) Typical evolution of the droplet during detachment of a granular suspension ($\phi=6\%$, $d_{part}=80\mum$). $W_{min}$ is the minimal neck radius, $L$ the length between the nozzle and the bottom of the droplet and $t_p$ represents the moment of final detachment, the time of pinch. The box in the bottom left is $1\milli\meter \times 1\milli\meter$.}
\end{center}
\end{figure}

The thinning dynamics of the neck is typically described by the evolution of the minimum neck width $W_{min}$ with time to pinch off $t_p-t$ (see \fig{fig:detach_susp}). \fig{fig_regimes_POF} shows results from our earlier work \citep{Bonnoit2010Accelerated} and illustrates the different regimes. At the beginning of the detachment, corresponding to the right side of the graph, the thinning dynamics depends on volume fraction, corresponding to the effective fluid regime. Then all curves collapse onto a single curve, identical to the interstitial fluid, representing the interstitial fluid regime. Finally the suspensions detach more quickly compared to the interstitial fluid (accelerated regime).
While it is commonly admitted that at the beginning of the detachment the dynamics are entirely given by the effective shear viscosity of the suspension \citep{Bonnoit2010Accelerated, Bertrand2011Dynamics, furbank2007pendant} there is still debate about the exact form and origin of the detachment dynamics close to the final pinch off \citep{Bonnoit2010Accelerated, Miskin2011DropletA}.

The existence of different detachment regimes and the continuous crossover between these regimes makes the analysis of the detachment dynamics complex. To address some of the still open questions, we here extend our previous work to very low volume fractions. At these small volume fractions the viscosity of the suspension is hardly different from the viscosity of the interstitial fluid.  In this case, the effective and interstitial fluid regime are both identical to the thinning dynamics of the pure interstitial fluid and we can directly study the influence of particles on the accelerated regime.

\section{Setup and data analysis}
\label{sec:setup}

For our measurements, we have created monodisperse suspensions of spherical polystyrene particles (Microbeads Dynoseeds) of diameter $d_{\textrm{part}}=20, 40, 80,$ and $140 \micro\meter$ in Shin-Etsu KF-6011 silicone oil. The oil's measured shear viscosity at  T=$22\celsius$  is $\eta_{\textrm{int}}=180 \milli\pascalsecond$; its surface tension is $\sigma=21\milli\newton\per\meter$. The density of both the particles and the oil is $\rho=1.08\gram\per\milli\liter$, resulting in a neutrally buoyant suspension. The size of the particles and thetime scales considered are such that Brownian motion can be neglected.

The shear viscosity of our suspension in bulk can be well described using the \cite{zarraga2000characterization} model:
$$\eta_{\textrm{eff}} = \eta_{\textrm{int}}\frac{\exp(-2.34\phi)}{(1-\phi/\phi_m)^3}, \phi_m\approx 0.62$$
where $\eta_{\textrm{int}}$ is the viscosity of the fluid between the particles \citep{bonnoit2010inclined}.

\begin{figure*}
\subfigure[\label{rmin low phi}]{\includegraphics[height=0.8\columnwidth]{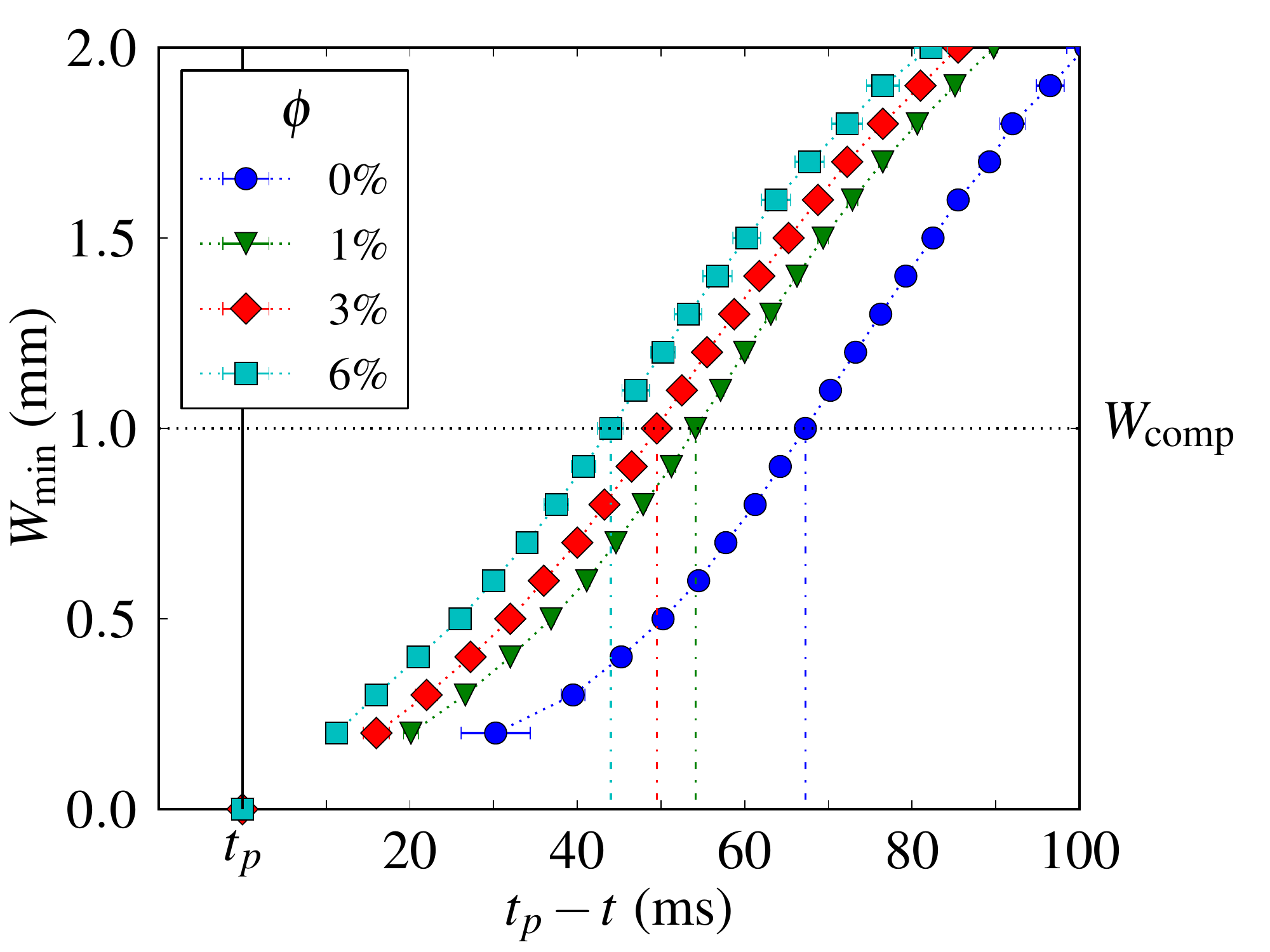}}
\hfill
\subfigure[\label{r_min_low_phi_shifted}]{\includegraphics[height=0.8\columnwidth]{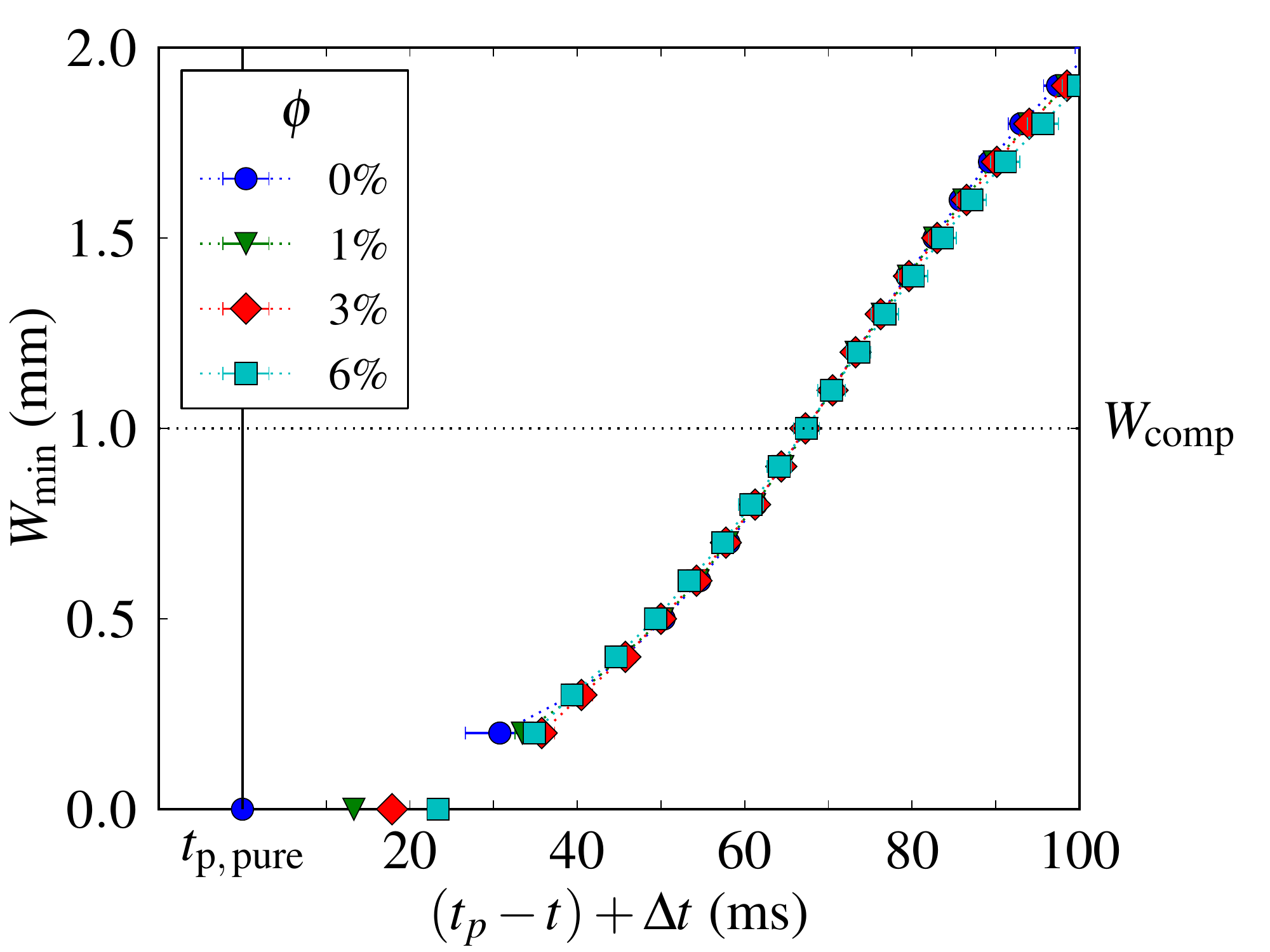}}
\caption{(a) $W_\text{min}(t_p-t)$ curves for low volume fraction suspensions for a grain diameter of $d_{part}=80\mum$. Shown are the averages and error bars in time over three experiments per suspension. Using $W_\text{comp}=1\ \milli\meter$, the determination of $\Delta t$ is illustrated. The relatively large change from the pure oil to the $1\%$ suspension compared to that of the $1\%$ to $3\%$ and $6\%$ is clearly visible. (b) The same curves, but shifted in time towards the pure oil to get a collapse at $W_{\textrm{comp}}$. The error bars are the same as in (a).\label{fig:curves_phi}}
\end{figure*}

We have used suspensions with a range of volume fraction $\phi=V_p/V=M_p/M$, where $V_p$ is the volume occupied by particles, and $V$ the total volume. Because the suspension is neutrally buoyant, this is equivalent to the ratio of the mass of the particles $M_p$ to the total mass $M$. For \sect{sec:low phi}, we have used $\phi={1,2,3,6}\%$. In \sect{sec:single particle}, our aim is to observe isolated particles in the neck and we have decreased the volume fraction to $5000$ particles per $\milli\liter$, corresponding to $\phi=0.12\%$ for $d_{\textrm{part}}=80\micro\meter$ and $\phi=0.53\%$ for $d_{\textrm{part}}=140\micro\meter$. With this volume fraction individual particles get trapped in the thread.

The suspensions are extruded from a $1\milli\liter$ syringe, with a standard Luer nozzle ($4\milli\meter$ external diameter), by a syringe pump, at an average rate $Q=53\micro\liter\per\minute$, corresponding to a few drops per minute. The experiments are performed at room temperature (T=$22\celsius$). The droplets are imaged at $4000\mathrm{fps}$ using a Photron Fastcam SA-3 high-speed camera with a Sigma 105 mm F/2.8 EX DG macro lens. This allowed us to image with a resolution of up to $1024 \times 256\mathrm{px}$ at $30\micro\meter\per\mathrm{px}$.
To improve contrast, we used a high-power LED light source to light the droplets from behind. Using an edge detection algorithm with linear interpolation we could then determine the neck width $W(z,t)$.
As in \cite{rothert2003formation}, we fit a parabola to $\pm 10\textrm{ pixels}$ around the minimum of $W(z)$ to determine the minimum neck diameter $W_{\textrm{min}}(t)$. Using this method we obtain reliable data for values of the minimal neck diameter larger then 200$\mum$. The minimal neck diameter is then represented as a function of $t_p-t$, with $t_p$ the time of pinch.

To determine the time of pinch $t_p$, we manually determine the first frame after detachment, and use that frame as the zero point in time. In this way, even if in the last stages of the detachment the minimal neck diameter is outside our experimental resolution, the time of pinch is determined with high precision corresponding to the time interval between two frames 0.25~ms. For each experiment, we determine $t_p-t$ for $W_{\textrm{min}} = 0.2\ldots 2\milli\meter$, in $0.1\milli\meter$ steps. For each volume fraction $\phi$, we then show the average $t_p-t$ and the standard deviation as error bars.

We have also measured $L$, the distance between the syringe and the bottom of the droplet, by manually measuring $L_{\textrm{syringe}}$, and by determining $L_{\textrm{bottom}}$ at the pinch off.

\section{Low volume fraction suspensions}
 \label{sec:low phi}

\begin{figure*}
\begin{minipage}[b]{0.45\linewidth}
\centering
\subfigure[\label{fig:figure_4c}]{\includegraphics[width=0.8\linewidth]{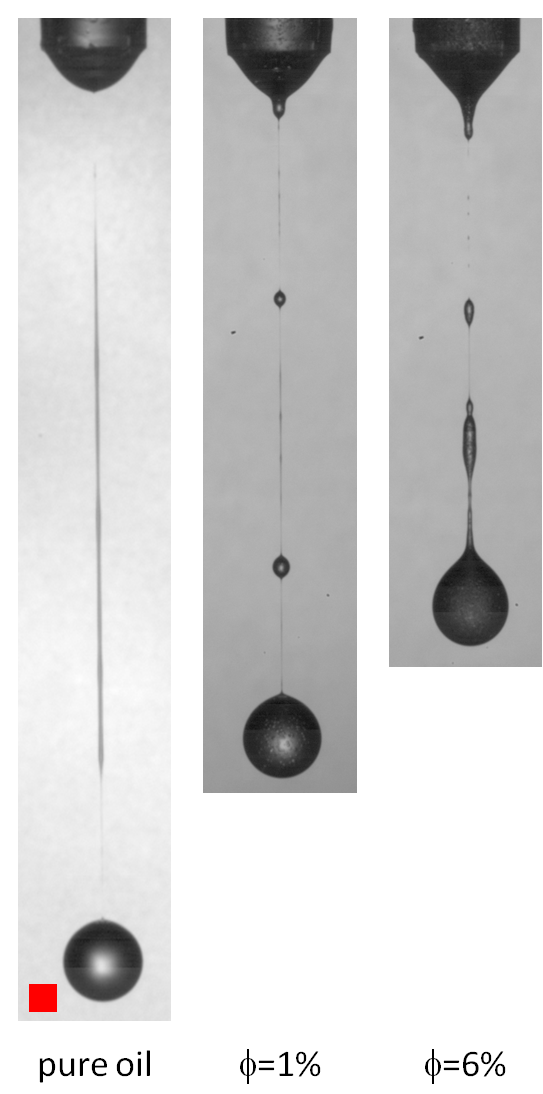}}
\end{minipage}
\begin{minipage}[b]{0.55\linewidth}
\centering
\subfigure[\label{fig:figure_4a}]{\includegraphics[width=0.8\linewidth]{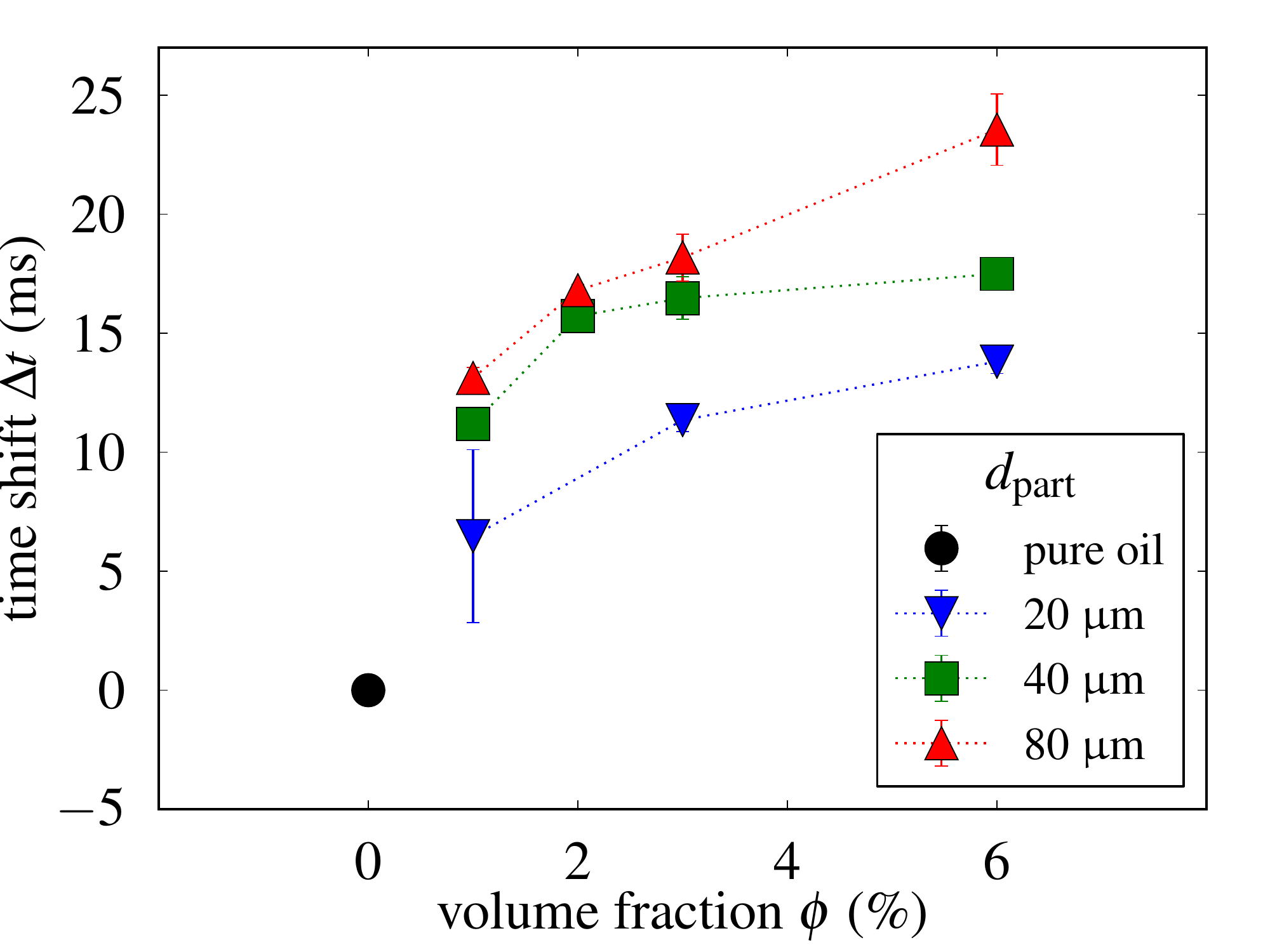}}\\
\subfigure[\label{fig:figure_4b}]{\includegraphics[width=0.8\linewidth]{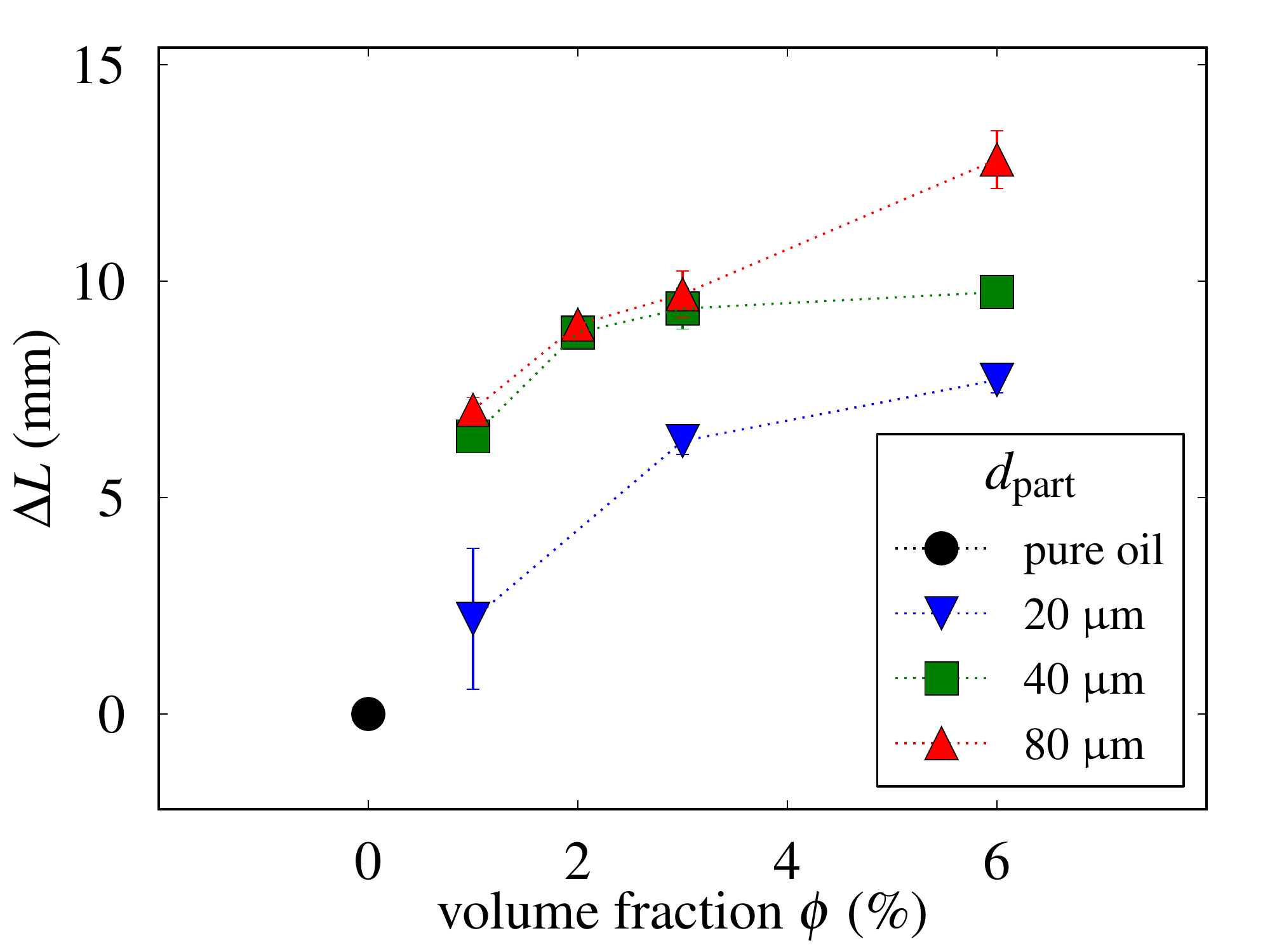}}
\end{minipage}
\caption{(a) Pinch frames for the interstitial fluid and a $1\%$ and a $6\%$ suspension of $80 \micro\meter$ particles. The structures in the suspension threads are apparent. The box in the bottom left is $1\milli\meter \times 1\milli\meter$.
(b) and (c) Time shift $\Delta T$ and difference in thread length $\Delta L$ between pure oil and suspensions of $\phi={1,3,6}\%$ and $d_{\textrm{part}} = \{20,40,80\} \micro\meter$. Data points represent averages over three experiments and the corresponding standard deviation. Data from \fig{rmin low phi} for $d_{\textrm{part}} = 80\micro\meter$ and from additional experiments for particle sizes ($d_{\textrm{part}} = \{20,40\} \micro\meter$).
\label{fig:figure_4}}
\end{figure*}

In \fig{rmin low phi}, the thinning behavior is shown for the pure interstitial oil (blue circles), as well as for three of the four suspensions $\phi={1,3,6}\%$ for $d_{\textrm{part}} = 80 \micro\meter$. The minimal neck diameter is represented as a function of $t_p-t$, averaged in time over three experiments for each fluid. The data outside our spatial resolution ($W_{\textrm{min}}<200\mum$) is not shown. Note that our resolution in time allows to precisely determine the time of pinch $t_p$ for each experiment even if we cannot  spatially resolve the thinning dynamics close to the pinch off.

When comparing the time to final detachment of the droplets at a given minimal neck width $W_{\textrm{comp}}$, one can see that the suspensions detach more quickly than pure oil. The detachment is thus accelerated for the suspensions and this acceleration increases with increasing volume fraction.

To get more insight in the thinning dynamics of the suspensions compared to the pure oil (\fig{r_min_low_phi_shifted}), we have shifted the curves in time by $\Delta t$ to obtain data collapse at $W_{\textrm{comp}}=1\milli\meter$  (different values for $W_{\textrm{comp}}$ give identical results). The time evolution of the neck width is similar for the pure oil and the suspensions away from the final pinch off. This is due to the fact that for the low volume fractions used in this study, the viscosity of the suspensions stays very close to the viscosity of the suspending fluid, $\eta_{\textrm{eff}} \approx \eta_{\textrm{int}}$. Based on the \cite{zarraga2000characterization} model, we estimate $\eta_{\textrm{eff}} = 1.17 \eta_{\textrm{int}}$ for a $6\%$ suspension, while for a $1\%$ suspension, $\eta_{\textrm{eff}} = 1.03 \eta_{\textrm{int}}$. Such small differences in viscosity induce a very small change in the thinning dynamics, which fall outside our experimental resolution.

The acceleration of the thinning takes place close to the final pinch off and is found outside our spatial resolution. From the time evolution of the neck radius it is thus difficult to determine the exact moment when the thinning dynamics of the suspensions start to differ from pure oil. The acceleration of the detachment however has a clear influence on the length of the filament at detachment, as can be seen from the snapshots shown in \fig{fig:figure_4c}. The filament is significantly shorter for the suspensions.

We have thus quantified the effect of the acceleration by measuring on the one hand the time shift $\Delta t$, visible through the markers on the x-axis representing the pinch off for the suspensions. A large $\Delta t$ corresponds to a strong acceleration of the detachment. \fig{fig:figure_4a} shows $\Delta t$ for different volume fractions and grain diameters. A strong increase of the time shift is observed when going from pure fluid to a small volume fraction. When further increasing the volume fraction the time shift continues to increase. Larger particles lead to a stronger acceleration at identical volume fraction. The acceleration of the detachment thus increases with volume fraction and grain diameter. On the other hand the change in thread length between the suspensions and the pure oil $\Delta L$, with $L$ the distance between the syringe and the bottom of the droplet at $t=t_p$, is quantified on \fig{fig:figure_4b}. $\Delta L$ increases with volume fraction and grain diameter. Comparing \fig{fig:figure_4a} and \ref{fig:figure_4b} it is obvious that $\Delta t$ and $\Delta L$ are directly linked as the dependence on volume fraction and grain size is observed to be identical for both quantities. Both $\Delta t$ and $\Delta L$ might be given by a critical neck width at which the thinning dynamics start to deviate from the thinning of the filament of pure oil. In \cite{Bonnoit2010Accelerated} we have shown that for higher volume fractions this critical neck width is indeed a function of the grain size and of the volume fraction, in agreement with the present observations.

We have shown in this section that for suspensions of small volume fraction the thinning dynamics away from the pinch off are identical to the thinning of pure oil. As a consequence the effective and interstitial fluid regimes, observed for dense suspensions, are identical. Using suspensions of very small volume fraction thus allowed us to investigate directly the role of the particles on the accelerated regime and our experiments unambiguously show that the presence of particles leads to an accelerated detachment of dilute suspensions compared to the pure interstitial fluid.

\section{Single particles in the thread}
\label{sec:single particle}

\begin{figure*}
\hfill\subfigure[$t=t_p-15\milli\second$]{\includegraphics[width=0.3\columnwidth]{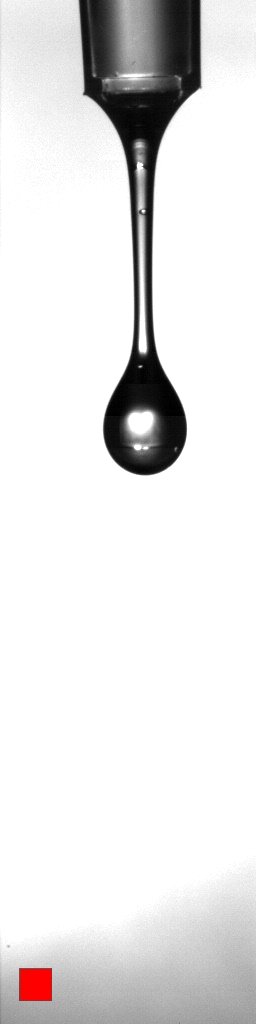}}
\hfill\subfigure[$t=t_p-11.7\milli\second$]{\includegraphics[width=0.3\columnwidth]{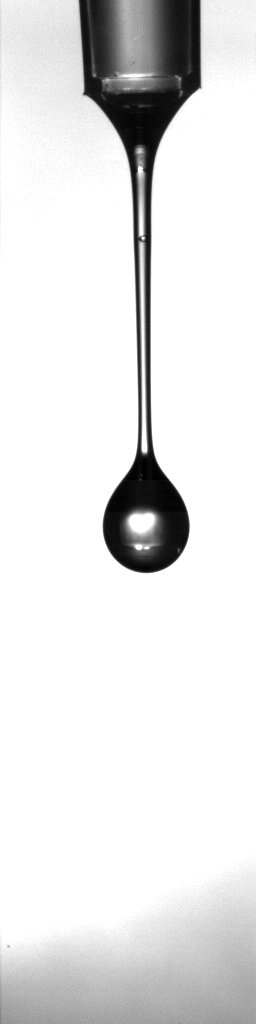}}
\hfill\subfigure[$t=t_p-8.3\milli\second$]{\includegraphics[width=0.3\columnwidth]{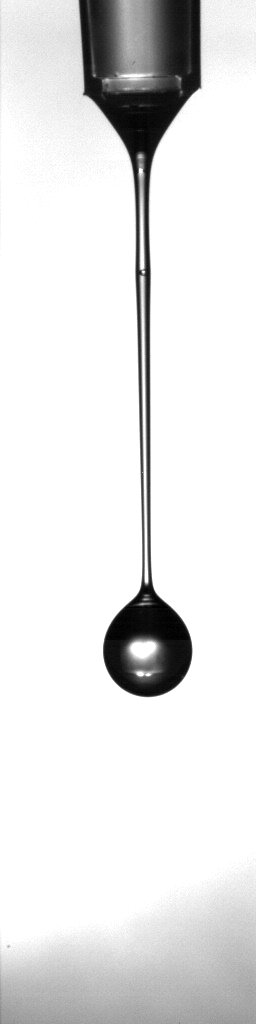}}
\hfill\subfigure[$t=t_p-5\milli\second$]{\includegraphics[width=0.3\columnwidth]{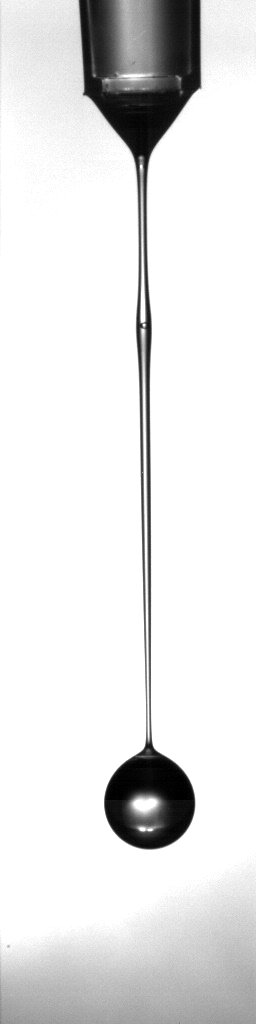}}
\hfill\subfigure[$t=t_p-1.7\milli\second$]{\includegraphics[width=0.3\columnwidth]{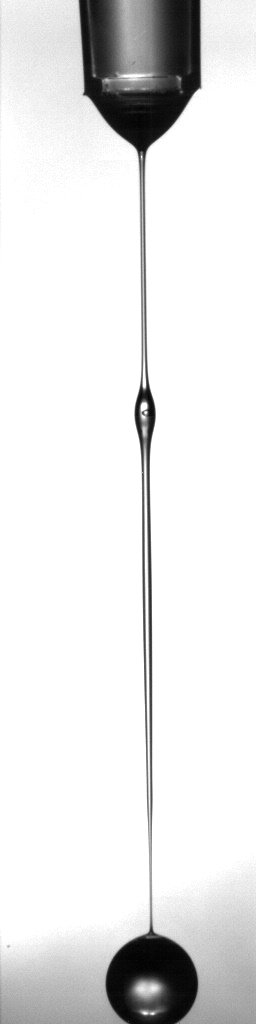}}
\hfill\subfigure[$t=t_p$]{\includegraphics[width=0.3\columnwidth]{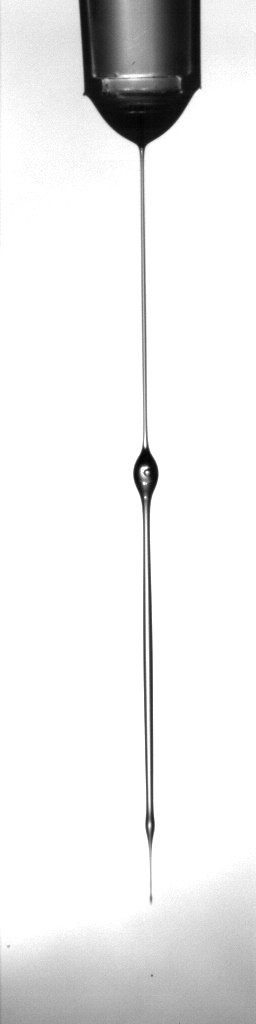}}
\hfill \,
\caption{Example of the effect of a particle on the droplet detachment. For illustrative reasons pictures using a large particle ($d_{part}=250\mum$) has been used. The box in the bottom left is $1\milli\meter \times 1\milli\meter$.
 \label{fig:nhung_snapshots}}
\end{figure*}

In this section we try to get more insight into the origin of the acceleration of the detachment observed for low volume fractions by
investigating the role of individual particles trapped in the thread. To do so, we have extended our study even further into the dilute regime and work at very low volume fractions (see \sect{sec:setup}) where individual particles get captured in the thread. \fig{fig:nhung_snapshots} illustrates the strong modification of the detachment caused by the presence of a single particle. To illustrate the phenomenon more clearly a very large particle ($d_{\textrm{part}}=250\mum$) is shown, but identical observations are made for smaller particles. The particle starts to modify the thinning of the thread when the neck diameter is still larger than the particle size and the formation of a large drop of fluid around the particle is observed. This has also been observed by \cite{furbank2004experimental} and generates profiles comparable to wetting drops on a fiber  \citep{carroll1984equilibriumliquiddrops}. The particle "pumps" fluid out of the thread because of the Laplace pressure difference between both and separates in this way the thinning thread into two separate filaments. The particle sets in this way a new scale at the boundary between the two filaments  replacing the scale set by the nozzle for the initial filament. These new and shorter filaments then continue to thin independently until one of the two filaments causes the final pinch off.

\begin{figure*}
\subfigure[\label{fig:singleparticlesamples}]{\includegraphics[height=0.8\columnwidth]{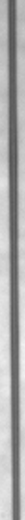}\hspace{0.05\columnwidth}
\includegraphics[height=0.8\columnwidth]{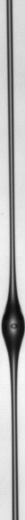}\hspace{0.05\columnwidth}
\includegraphics[height=0.8\columnwidth]{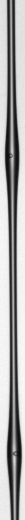}}
\hfill
\subfigure[\label{fig:boxplot}]{\includegraphics[height=0.8\columnwidth]{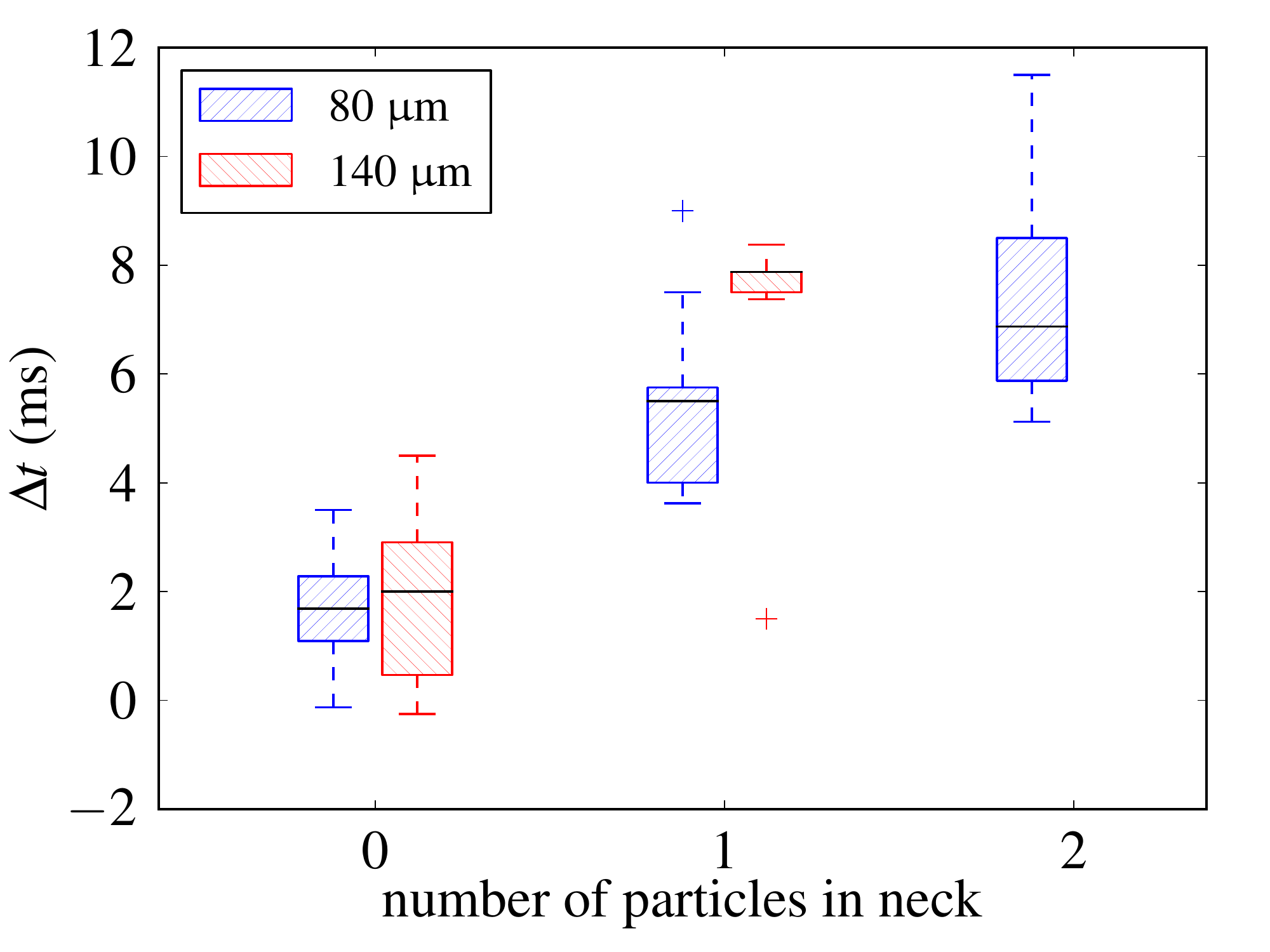}}\hfill
\subfigure[\label{fig:nhung_bubbles}]{\includegraphics[height=0.8\columnwidth]{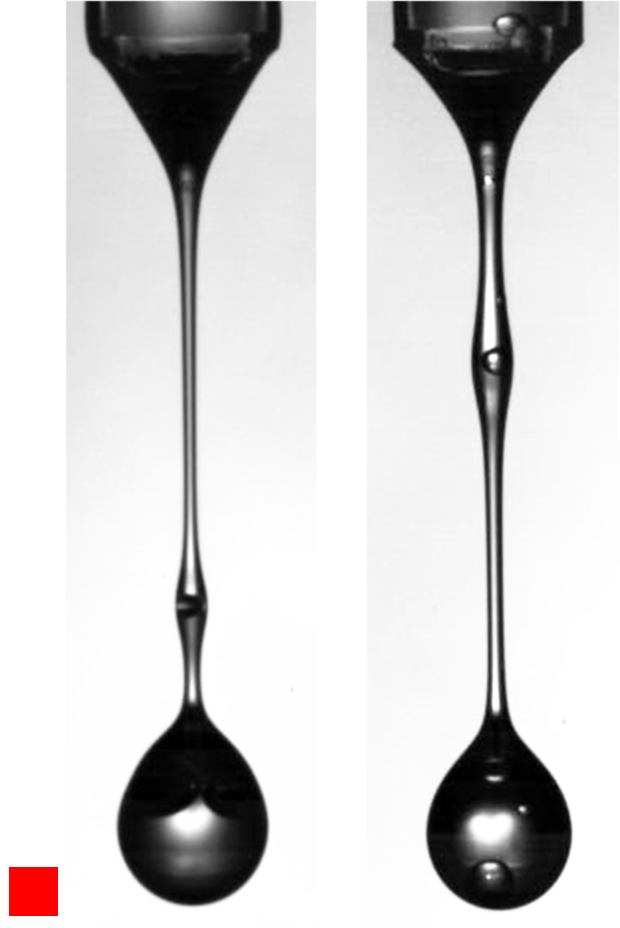}}
\caption{(a) Snapshots of necks with zero, one and two particles. The snapshots are $1\milli\meter \times 9.5\milli\meter$. (b) Boxplot of the shifts for the different numbers of particles in the neck. Each box contains more then 10 experiments. The box indicates the $25-75\%$ percentile, while the whiskers indicate the furthest value inside 1.5 IQR. Outliers are shown indicated by $+$ symbols. ($d_{\textrm{part}}=80\mum$) (c) Comparison between a bubble (left) and a particle of $d_{\textrm{part}}=500\mum$ (right) captured in the thread. The box is $1\milli\meter \times 1\milli\meter$.}
\end{figure*}

We classify our experiments, using visual observation, into detachments with none, a single or two particles in the neck (see \fig{fig:singleparticlesamples}). The rare detachments with more than two particles in the neck are ignored. We will analyze our results as a function of these three cases and compare them to pure oil.

Qualitatively, the thinning behavior $W_{\textrm{min}}(t_p-t)$ for the three cases (not shown) is found to be identical to the observations made for the dilute suspensions: away from the pinch off the thinning is observed to be identical to the pure fluid, but close to final pinch off the detachment is accelerated.  Note that the case with no particle in the thread also differs slightly from the experiments performed with the pure fluid. This is most likely due to the fact that even without particles in the thread, a particle might still be close to the junction of the thread with the droplet where it might have a small influence on the detachment. We have quantified the acceleration by measuring the time shift for the different cases shown in \fig{fig:boxplot}. The acceleration is found to increase with increasing number of particles and particle size in line with the observations made for the dilute suspensions.

The time shift and thus the acceleration observed even for a single particle is significant (\fig{fig:boxplot}) and corresponds to 50$\%$ of the time shift observed for a suspension of $\phi=1\%$ (\fig{fig:figure_4a}). The presence of a single particle in the thread thus strongly modifies the detachment of a droplet. The separation of the thinning filament into two shorter and thinner filaments, induced by the particle, might be at the origin of the observed acceleration of the detachment. Similar mechanisms are likely to be at the origin of the accelerated detachment of the dilute suspensions. To establish a more clear link between the measurements with single particles and the dilute suspensions, statistics on the distribution of particles in the thread would be needed for the dilute suspensions. The process of capturing individual particles in the thread results from an interaction between the particle and the local flow field in the thread \citep{gier2012visualizationflowprofile} and \cite{furbank2004experimental} have shown that it depends on the volume fraction, particle size and the nozzle radius. But so far no detailed study of this process exists.

In a theoretical paper \citet{hameed2009breakup} have studied the effect of a single particle on droplet detachment and have found a stabilizing effect of the particle leading to a more slow thinning of the filament. This is in contrast to our experimental findings. In the model the particle is held at a fixed position whereas it is free to move in the thread in the experiments. This might be at the origin of the differences between the predictions of the model and the experimental observation. Further investigations are needed to understand the effect of the particle on the thinning dynamics in detail.

In addition to the experiments performed with particles we have also trapped bubbles in the thread. Interestingly a bubble and a particle in the thread lead to exactly the same thinning dynamics (data not shown) as is illustrated on the snapshots shown in \fig{fig:nhung_bubbles}.

\section{Particle tracking}
\label{sec:PIV}

\begin{figure}
\subfigure[$t_p-26.25\milli\second$ \label{fig:piv_1}]{\includegraphics[width=0.9\columnwidth]{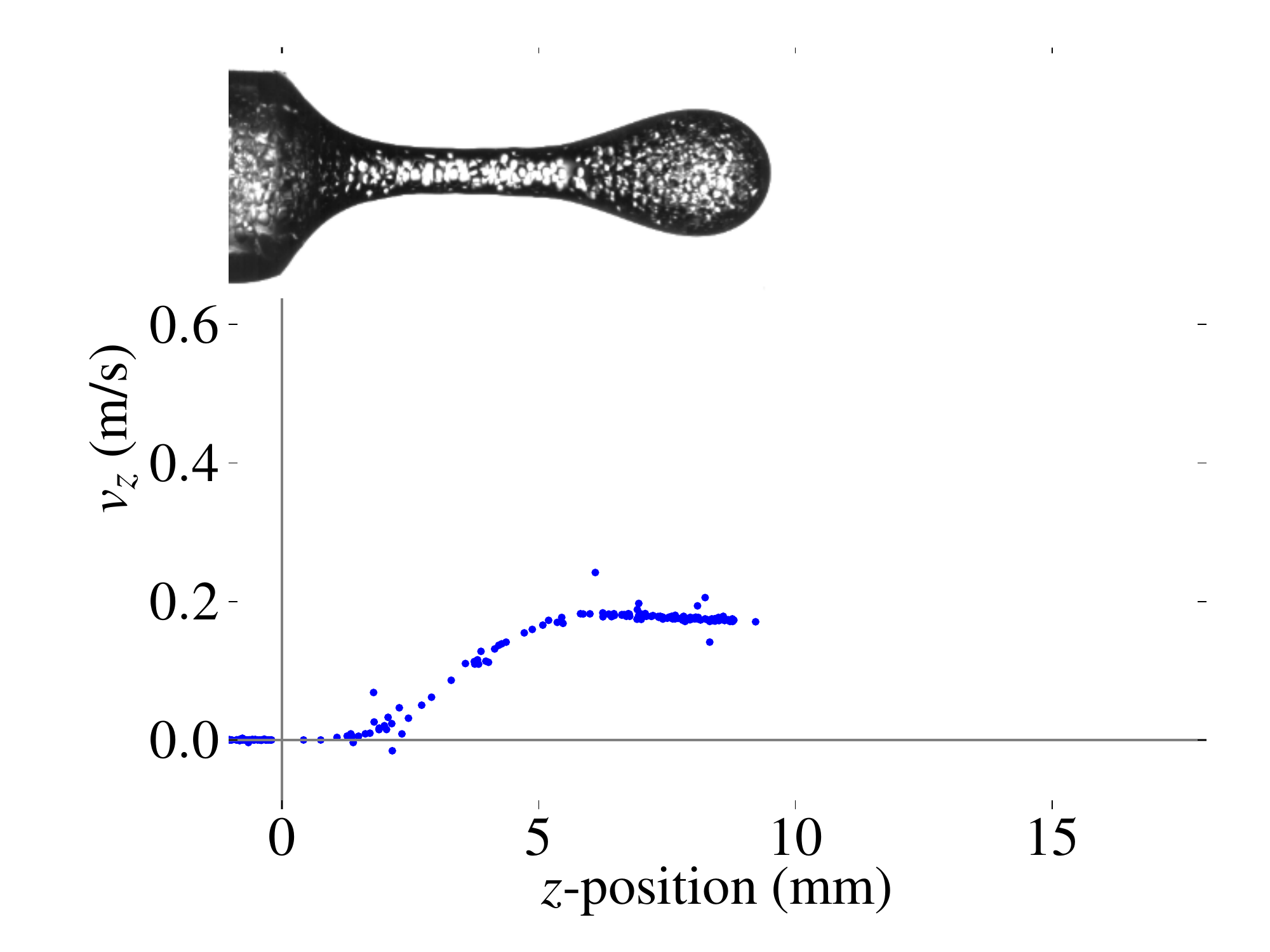}}
\hfill
\subfigure[$t_p-8.5\milli\second$ \label{fig:piv_2}]{\includegraphics[width=0.9\columnwidth]{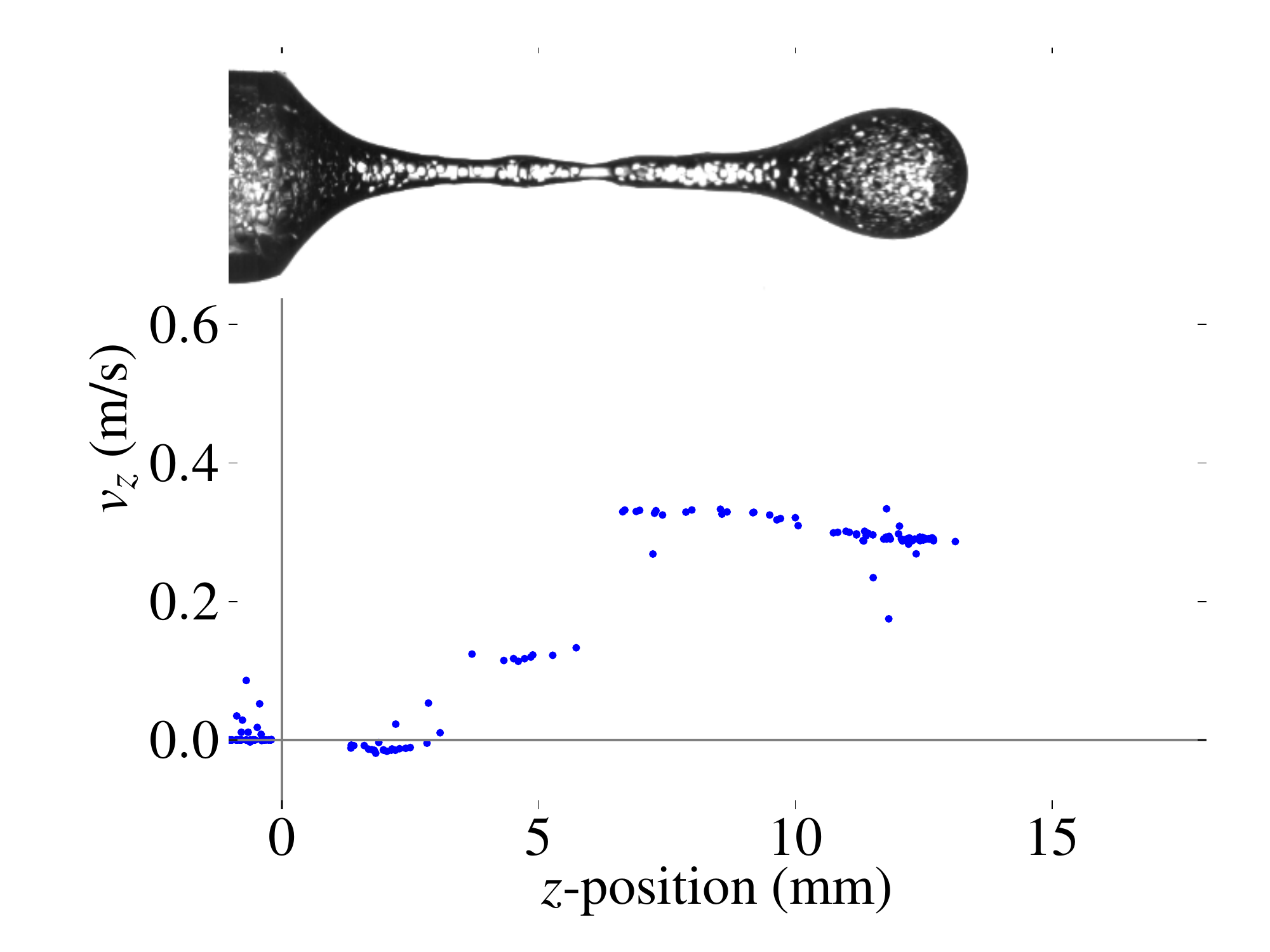}}
\hfill
\subfigure[$t_p-2\milli\second$\label{fig:piv_3}]{\includegraphics[width=0.9\columnwidth]{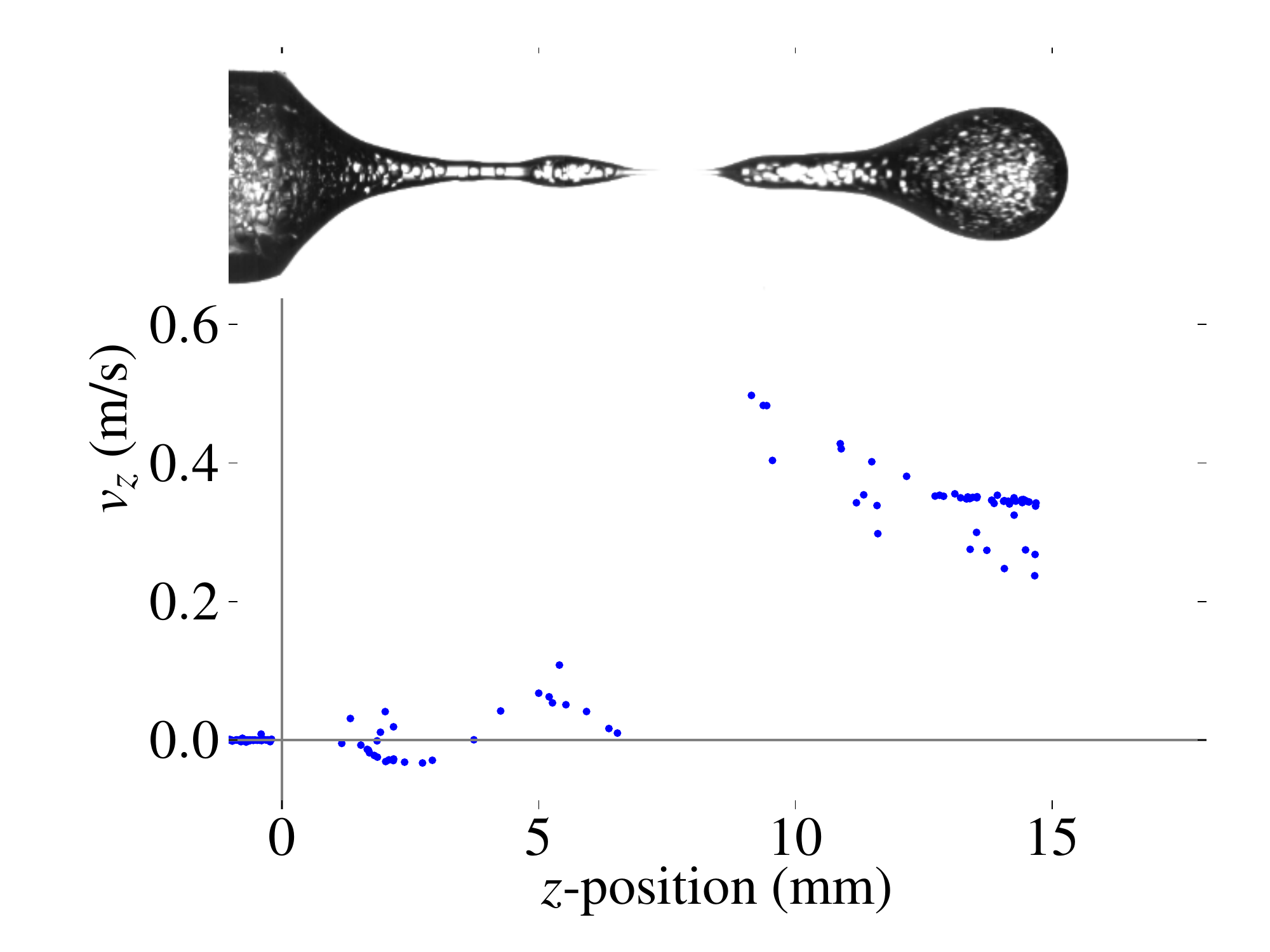}}
\hfill\,
\newline
\caption{PIV graph, showing localized thinning. The scale of the figure corresponds to that on the x-axis. In (b), two thinning regions can be seen: around $z=3.5\milli\meter$ and $z=6.5\milli\meter$, which can be recognized by the jump in $dz/dt$. \label{fig:piv}}
\end{figure}

To link our observations on suspensions of very small volume fractions to the detachment of more dense suspensions, we have analyzed the thinning dynamics for a $\phi=15\%$ suspension with large ($d=250\micro\meter$) particles using PIV. In \fig{fig:piv}, we show snapshots and the corresponding velocities of particles in the direction of the axis of the thread, averaged over the thread radius. The velocity of a particle is determined using a particle tracking algorithm (MTrack2) and smoothed over $2.5\milli\second$. In \fig{fig:piv_1} one can observe that the part of fluid close to the nozzle is at rest whereas the droplet moves at a constant velocity. The velocity increases continuously between these two areas indicating a continuous thinning of the thread. In \fig{fig:piv_2} the apparition of another area with a constant velocity is observed. This corresponds to the formation of a thicker part of the thread which is not deformed any more and the thinning becomes localized above and below this structure. In \fig{fig:piv_3} thinning is entirely localized in the part of the thread left of the structure. These results indicate that during detachment of suspensions thinning of the thread becomes localized and very soon only small parts of the thread continue to thin.  We think that this localization is caused by rearrangements of particles freeing areas of grains. In these areas the viscosity is deceased leading to an accelerated and subsequently localized thinning. These areas correspond to threads of suspensions of small volume fraction. In the present paper we have shown that such threads thin similarly to the interstitial fluid and that towards the end of the detachment the presence of particles leads to an acceleration. Our current study thus also quantifies the transition between the interstitial fluid regime and the accelerated regime observed in more dense suspensions (see \fig{fig_regimes_POF}).

Note that localized thinning of a thread has also been observed for a shear thickening suspension  \citep{Roche2011Heterogeneity} and has been attributed to a partial jamming of the suspension. In our case the localization is also due to a strong viscosity contrast between the areas less or more dense in grains. The existence of a jammed state is however unlikely in our system as localization of the thinning is observed for average volume fraction as small as $\phi=15\%$ as shown here.

\section{Conclusion}
\label{sec:conclusion}

We have shown that the presence of a single particle in the viscous thread linking a drop to a nozzle leads to a significant change in the thinning dynamics: the detachment is accelerated. We have quantified this acceleration for different numbers of particles captured in the thread and for different particle sizes. We have established the link between the presence of individual particles in the thread and observations made for suspensions of small volume fraction, where several particles get trapped. The dynamics of capturing particles in the thread is far from trivial \citep{furbank2004experimental} and not fully understood yet. The acceleration observed for suspensions of small volume fractions results from the probability of capturing given numbers of particles in the thread.

We have also shown that our results for small volume fractions might be useful to understand the detachment dynamics of suspension of higher volume fractions. For dense suspensions particles rearrange during the thinning of the thread and form areas less dense in particles. The thinning soon localizes at these areas that contain less particles and in this way also represent suspensions of small volume fraction. Our study on the detachment of suspensions of very small volume fractions thus explains the accelerated detachment observed in previous studies \citep{furbank2007pendant, Bonnoit2010Accelerated}. In the future it might be interesting to study the exact dynamics of rearrangement of particles in the thread and thus clarify the picture of the link between the role of individual particles and the thinning dynamics of dense suspensions.

\begin{acknowledgements}
MvD would like to thank the French embassy in The Netherlands for
their funding under the Bourse d'excellence \textquotedbl{}Descartes\textquotedbl{}
program. We thank Martin van Hecke and Marc Miskin for interesting discussions and P.-B. Bintein for help with the experimental set-up.
\end{acknowledgements}
\bibliographystyle{spbasic}
\bibliography{citedrefs}

\end{document}